\documentclass[imes,tighten,twocolumn,twocolappendix]{aastex7}

\newcommand{\aafrag}{{\tt AAfrag}}
\newcommand{\aafragv}{{\tt AAfrag202}} 
\newcommand{\GP}{{\scshape GALPROP}}
\newcommand{\icecube}{IceCube}
\newcommand{\hess}{{H.E.S.S.}}

\newcommand{\lhaaso}{LHAASO}
\newcommand{\fermi}{\textit{Fermi}--LAT}
\newcommand{\graya}{$\gamma$-ray}  
\newcommand{\grays}{$\gamma$-rays} 
\newcommand{\hi}{H\,{\scshape i}}
\newcommand{\htwo}{H$_2$}

\shorttitle{Diffuse Neutrino Emission from the Milky Way}
\shortauthors{Marinos et al.}

\makeatletter
\newcounter{paperfn}
\renewcommand\thepaperfn{\arabic{paperfn}}
\newcommand\paperfootnote[1]{%
  \refstepcounter{paperfn}
  \protected@xdef\@thefnmark{\thepaperfn}
  \@footnotemark
  \@footnotetext{#1}
}
\makeatother

\begin{document}

\title{Simulating the Diffuse Neutrino Emission from the Milky Way with \GP{}}

\author[0000-0003-1734-0215,sname='Marinos']{P. D. Marinos}
\affiliation{W. W. Hansen Experimental Physics Laboratory and Kavli Institute for Particle Astrophysics and Cosmology, Stanford University, Stanford, CA 94305, USA}
\email[show]{pmarinos@stanford.edu}

\author[0000-0002-2621-4440,sname='Porter']{T. A. Porter}
\affiliation{W. W. Hansen Experimental Physics Laboratory and Kavli Institute for Particle Astrophysics and Cosmology, Stanford University, Stanford, CA 94305, USA}
\email{tporter@stanford.edu}

\author[0000-0002-9516-1581,sname='Rowell']{G. P. Rowell}
\affiliation{School of Physical Sciences, University of Adelaide, Adelaide, South Australia 5000, Australia}
\email{gavin.rowell@adelaide.edu.au}

\author[0000-0001-6141-458X,sname='Moskalenko']{I. V. Moskalenko}
\affiliation{W. W. Hansen Experimental Physics Laboratory and Kavli Institute for Particle Astrophysics and Cosmology, Stanford University, Stanford, CA 94305, USA}
\email{imos@stanford.edu}

\author[0000-0003-1458-7036,sname='Jóhannesson']{G. Jóhannesson}
\affiliation{Science Institute, University of Iceland, IS-107 Reykjavik, Iceland}
\email{gudlaugu@hi.is}

\begin{abstract}

    We use the \GP{} cosmic ray (CR) propagation framework to model the diffuse neutrino and gamma-ray (\graya{}) emissions from the Galaxy.
    A collection of realistic bounding models are developed and predictions of the resulting neutrino and \graya{} signals are compared to the \icecube{} and \lhaaso{} data up to PeV energies.
    We find that all the \GP{} models are consistent with the neutrino data within uncertainties.
    They are also consistent with expectations of neutrino emissions derived from \lhaaso{} data when accounting for possible \graya{} point source contamination.
    The new models present state-of-the-art predictions for the VHE neutrino emissions from the Galaxy that may be used for future neutrino searches. 

\end{abstract}




\section{Introduction}

Cosmic rays (CRs) diffuse throughout the Milky Way (MW), interacting with the interstellar medium (ISM) and creating secondary CRs, gamma rays (\grays{}), and neutrinos.
These CRs and \grays{} have been observed and studied for decades across many orders of magnitude in energy.
However, the CR propagation and injection mechanisms, as well as the hadronic/leptonic fraction, remain unconstrained.
With the recent neutrino detection of the Galactic plane by \icecube{} we have an additional multi-messenger dataset that can aid in constraining the propagation and injection of CRs within the MW.

\icecube{} is a cubic-kilometre detector of $>$1\,TeV neutrinos located at the South Pole.
Prior to 2023, \icecube{} had only detected two astrophysical sources of very-high-energy (VHE) neutrinos: the blazar TXS~0506+056 \citep{2018Sci...361..147I,2018Sci...361.1378I}, and the active galaxy NGC~1068 \citep{2022Sci...378..538I}.
The third and most recent detection of a VHE neutrino source was the Galactic plane \citep{2023Sci...380.1338I}.
However, \icecube{} was not able to detect the MW in a blind search, relying on input models to define the expected neutrino signal from the Galaxy.
The three input models used by \icecube{} are based on CR diffusion codes.
One input, the `$\pi^{0}$' model, is based on a 13-year-old \GP{} result \citep{2012ApJ...750....3A} tuned only to CR data.
The other two models are based on Kraichnan-diffusion models tuned to CR and \graya{} data (named `KRA$_{\gamma}$') calculated with the DRAGON CR propagation code \citep{2015ApJ...815L..25G}.
All three input spatial templates resulted in a successful detection of the MW plane, from which \icecube{} was able to measure the Galactic neutrino flux.
These neutrino flux measurements then represent the sum of the source and diffuse components.

While neutrinos have not yet been used as a proxy measurement for the Galactic CR density, the \graya{} emission has been used for decades in the MeV--GeV regime.
Various \graya{} observatories have reported a detection of the diffuse/large-scale emission in the TeV regime.
One of these, the \lhaaso{} observatory, detects \grays{} in the energy range of 0.5--500\,TeV.
The $pp$ collisions generating these \grays{} also create neutrinos with energies ($E_{\nu}$) in the range of 0.25--250\,TeV.
As the \lhaaso{} and \icecube{} energy ranges overlap, their observations are connected.
These neutrino and \graya{} datasets can then be correlated to one another, and can be used as upper-limits to constrain the hadronic components of Galactic \graya{} models.

In this paper we use up-to-date versions of the \GP{} framework\paperfootnote{\url{https://galprop.stanford.edu/}}, with the neutrino flux being computed with the latest interaction cross sections from \aafragv{}.
Using the steady-state (i.e.\@ time-independent) propagation models we compute the neutrino flux over a range of parameter configurations to provide a modelling uncertainty.
We also compute the uncertainties and related neutrino limits from the \lhaaso{} \graya{} observations and compare to our model predictions.
Our state-of-the-art models are then compared to the Galactic flux observed by \icecube{}.
We find that the \GP{} predictions lie within all upper and lower limits on the diffuse neutrino emission for both \lhaaso{} and \icecube{}.
We find that all of our model predictions are consistent with the range of current Galactic observations.
We also provide model neutrino predictions in the 1--100\,TeV energy range.
All of our model data products and configuration files are provided in the online material.

\section{Model Setup}

The \GP{} framework \citep{1998ApJ...493..694M,1998ApJ...509..212S} is a CR propagation package with an extensive history of reproducing the local CR spectra, Galactic synchrotron emission, and $\sim$keV to TeV \graya{} emission \citep[e.g.][]{2000ApJ...537..763S,2002ApJ...565..280M,2004ApJ...613..962S,2008ApJ...682..400P,2011A&A...534A..54S,2020ApJS..250...27B}.
Recently we have shown that models based on \GP{} version 57 \citep{2022ApJS..262...30P} can also reproduce the observed \graya{} emission in the TeV--PeV regime \citep{2023MNRAS.518.5036M,2025ApJ...981...93M} from the high-energy stereoscopic system (\hess{}) Galactic plane survey (HGPS) and the large high-altitude air shower observatory (\lhaaso{}).
In this paper we extend the models used in \citet{2023MNRAS.518.5036M, 2025ApJ...981...93M} to include the predicted neutrino emission.

We construct the ISM with multiple components: the interstellar radiation field (ISRF), the Galactic magnetic field (GMF), the gas distribution, and the source distribution (i.e.\@ where CRs are injected into the MW).
For all distributions, we chose models with spiral-arm descriptions of the MW.
For the ISRF we use the \citet{2012A&A...545A..39R} model (hereafter referred to as R12) as described by \citet{2017ApJ...846...67P}, and for the GMF we use the \citet{2011ApJ...738..192P} model (hereafter referred to as PBSS) as described by \citet{2013MNRAS.436.2127O}.
For the 3D molecular (\htwo{}) and neutral (\hi{}) gas densities we use the models developed by \citet{2018ApJ...856...45J}, and references therein.
Previous results from \citet{2023MNRAS.518.5036M} found that the hadronic emission has a strong dependence on the chosen source distribution.
To estimate uncertainty due to degeneracy in the source distributions we simulate over a range of models.
Hence, we simulate over a range of source distributions to estimate the variance between them.
We use distributions with disc-like and spiral-arm components \citep[see][and references therein]{2017ApJ...846...67P}.
The models are referred to by the percentage contribution from the spiral arm component -- e.g.\@ SA50 has an equal relative contribution from each component.
We simulate over three distributions: SA0, SA50, and SA100.
To replicate the diffuse emission calculation method used by the \lhaaso{} collaboration \citep{2023PhRvL.131o1001C} we also test a fourth method that takes the post-diffusion/propagated CR spectra at the Solar location from the SA50 model and applies it uniformly across the MW.
We hereby refer to this model as `u50'.

For the CR injection and diffusion parameters we take the values from \citet{2023MNRAS.518.5036M} (see their Table~1).
The parameters are obtained following the procedure in \citet{2017ApJ...846...67P} and \citet{2018ApJ...856...45J}.
For each source distribution (and ISM gas model) the propagated CR spectra are fit to data from AMS--02 and \textit{Voyager~1} \citep[see][and references therein]{2019ApJ...879...91J}.
The force-field approximation is applied instead of the more advanced \GP{}/HelMod framework as we are focusing on energies above which the Solar modulation has a significant impact \citep{2020ApJS..250...27B}.
The MW is constructed with a non-linear spatial grid~\citep[tan spatial grid;][]{2022ApJS..262...30P} with the spatial grid size around the solar location set to 7\,pc.
We use ten kinetic energy bins per decade ranging from 1\,GeV\,nuc$^{-1}$ to 10\,PeV\,nuc$^{-1}$ for the hadronic CR propagation, and five bins per decade ranging from 1\,GeV to 1\,PeV for the neutrino flux calculation.
The neutrino skymaps are computed on a seventh-order HEALPix~\citep{2005ApJ...622..759G} isopixelisation, giving a pixel size of $27.5^{\prime} \times 27.5^{\prime}$.
We simulate hadrons up to and including silicon.

CR observations in the TeV energy regime \citep[e.g.\@ DAMPE;][]{2019SciA....5.3793A} show a spectral `bump' (i.e.\@ a hardening followed by a softening).
This TeV bump is likely due to some local phenomenon, such as a nearby source or some reacceleration effect \citep[e.g.][]{2019JCAP...10..010L,2020ApJ...903...69F,2021ApJ...911..151M,2024AdSpR..74.4264M,2025ApJ...989...74B}.
However, the sharp breaks and a very thin transition between the excess in the outer Galaxy and a deficit in the inner Galaxy along the magnetic equator in the anisotropy map at 10\,TV \citep{2019ApJ...871...96A} rule out the likely sources beyond $\sim$10\,pc \citep{2024AdSpR..74.4264M}.

Previously, we performed the spectral tuning of the CRs at 100\,GeV and extrapolated into the TeV range to prevent overfitting to a potentially local effect.
Here we test over a range of CR injection indices to evaluate the sensitivity of the diffuse neutrino emission to the Galactic CR spectrum and to account for observational uncertainties in the $>$10\,TeV CR flux.
We take the previously used injection indices from \citet{2023MNRAS.518.5036M} for rigidities $R>266$\,GV for protons ($\gamma_{2,p} \sim 2.35$) and helium ($\gamma_{2,\mathrm{He}} \sim 2.3$) and vary both by $\pm 0.1$.
The \GP{} propagated CR proton and helium spectra are shown in the \hyperref[sect:appendix]{Appendix} in Figures \ref{fig:proton flux} and \ref{fig:helium flux}.
All our model parameter values are included within the configuration files supplied online.

\subsection{Neutrino Calculations}

We calculate the neutrino emissivity for CR proton and helium collisions with the hydrogen and helium gas, i.e.\@ $p$+$p$, He+$p$, $p$+He, and He+He interactions.
The yields for these interactions are computed with \aafragv{} \citep{2019CoPhC.24506846K,2023CoPhC.28708698K}.
While \aafrag{} also includes the yields for C+$p$, Al+$p$, and Fe+$p$ collisions, these heavier species have small relative CR and ISM abundances and so are not included.
We calculate the production of the electron and muon neutrinos and antineutrinos ($\nu_{e}$, $\nu_{\mu}$, $\bar{\nu}_{e}$, and $\bar{\nu}_{\mu}$), with the neutrino flux then being computed via a line-of-sight (LoS) integral over the emissivity.

As neutrinos propagate they oscillate between all three possible flavours: $\nu_{e}$, $\nu_{\mu}$, and $\nu_{\tau}$ \citep{1994hep.ph....8296L}.
Over parsec-scale distances the flux of each flavour is expected to be approximately equal \citep{2000PhRvD..62j3007A}, which has been confirmed in observational results from \icecube{} \cite{2015PhRvL.114q1102A,2025arXiv251024957A}.
The oscillations are not explicitly accounted for in the LoS calculations.
As the total all-flavour flux is not altered by the oscillations, we take the per-flavour emission as one third of the total.

\section{Results}

\begin{figure}
    \centering
    \includegraphics[width=\linewidth]{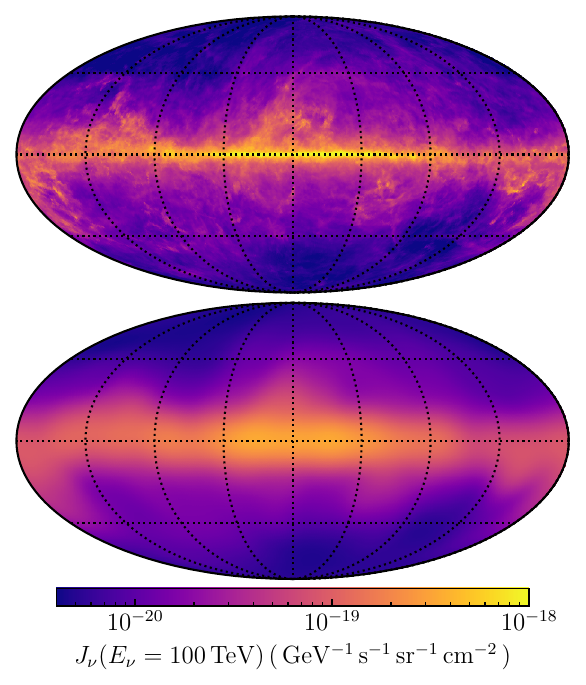}
    \caption{Skymap of the \GP{} predicted Galactic diffuse per-flavour neutrino flux at 100\,TeV. The top panel shows the \GP{} predictions for the SA50 source distribution, and the bottom panel is smeared with a $\sigma = 7^{\circ}$ Gaussian corresponding to the \icecube{} event uncertainty at 100\,TeV.}
    \label{fig:neutrino skymaps}
\end{figure}

\autoref{fig:neutrino skymaps} shows the \GP{} predicted Galactic diffuse per-flavour neutrino emission at the \icecube{} flux normalisation energy ($E_{\nu}=100$\,TeV).
In correspondence with the typical event directional uncertainty of \icecube{} at 100\,TeV we also show the \GP{} predictions with a Gaussian blur with a radius of $\sigma = 7^{\circ}$ \citep[][figure S5]{2023Sci...380.1338I}. 
The \icecube{} collaboration did not publish a skymap of their Galactic neutrino flux, only including a map of the all-sky pre-trial significance for a point-source search.
While there is a qualitative spatial agreement between our models and the \icecube{} neutrino event significance, a quantitative analysis would require using the \GP{} skymap predictions as an input in the \icecube{} analysis pipeline.

\begin{figure}
    \centering
    \includegraphics[width=\linewidth]{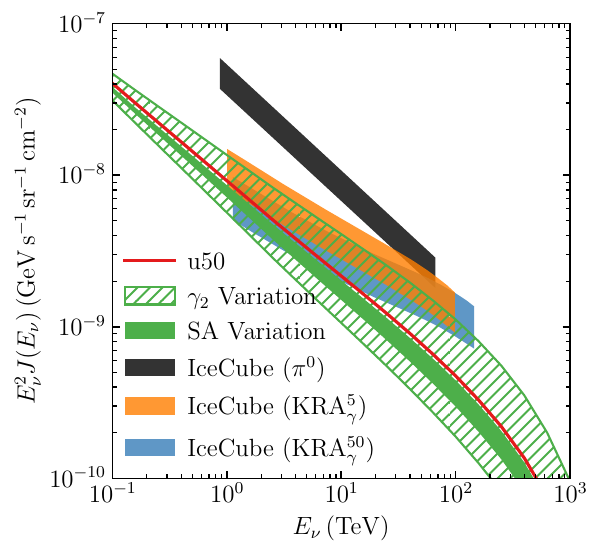}
    \caption{Envelopes of the \GP{} predicted all-sky diffuse per-flavour neutrino flux across the three source distributions (SA variation; green band) and over variations in the injection spectral index ($\gamma_{2}$ variation, green hatched band). Also shown are the three model-dependent \icecube{} Galactic plane flux measurements \citep{2023Sci...380.1338I}: $\pi^{0}$ (black), KRA$_{\gamma}^{5}$ (orange), and KRA$_{\gamma}^{50}$ (blue).}
    \label{fig:allsky neutrino flux}
\end{figure}

The \icecube{} detection of the MW plane used machine-learning algorithms to reduce the number of atmospheric neutrino events, increasing the statistics of astrophysical neutrinos by twenty times compared to their previous analysis in \citet{2019ApJ...886...12A}.
Despite this improvement in statistics, \icecube{} was unable to detect the MW in a blind search and instead relied on spatial models of the Galactic emission to extract the relevant neutrinos.
The three neutrino emission models utilised by \icecube{} are referred to as: $\pi^{0}$, KRA$_{\gamma}^{5}$, and KRA$_{\gamma}^{50}$, where the superscript on the KRA models refers to the cut-off energy of the proton spectrum.
These spatial templates are convolved with the detector acceptance and event angular uncertainties, giving event-specific spatial probability density functions that are then used in a maximum likelihood search.
If the Galactic plane is detected with the spatial template, then the model-dependent flux can be computed from the number of observed neutrino events.
These model-dependent fluxes take the spectral normalisation as a free parameter that depends on the number of signal events that are spatially correlated with the input models.
The \icecube{} $\pi^{0}$ model is normalised at $E_{\nu}=100$\,TeV.
As the KRA$_{\gamma}$ models have a more complex spectral shape and normalisation procedure, their normalisations are quoted only as multiples of their predicted flux \citep{2023Sci...380.1338I}.
The three model-dependent \icecube{} Galactic plane neutrino fluxes are shown in \autoref{fig:allsky neutrino flux} along with our computed range of \GP{} predictions.

We split our \GP{} results into three parts: the u50 flux, and envelopes over the three source distribtions (labelled SA variations) and the alterations to the CR injection spectral indices (labelled $\gamma_{2}$ variations).
For the changes in the SA models the spread in the envelope is largely due to minor differences in the spectral extrapolation above 1\,TeV.
While all three SA models and the u50 model have a similar spectral index to that of the \icecube{} $\pi^{0}$ model, we note that their analysis is not sensitive to the spectral shape of the neutrino emission.
The morphology of the Galactic neutrino emission is constant across energies 1--100\,TeV for all six of our models.

The \icecube{} collaboration was unable to resolve any neutrino point sources within the MW.
Their neutrino flux therefore represents the total neutrino emission from the Galaxy, i.e.\@ the sum of the diffuse and total source components.
The \icecube{} results in \autoref{fig:allsky neutrino flux} then show the large-scale neutrino emission, i.e.\@ the sum of source components and the diffuse emission.
The \icecube{} Galactic plane neutrino flux is then an upper limit on the hadronic diffuse flux.
We expect that the \GP{} predictions, which only includes contributions from the diffuse emission, should underestimate the \icecube{} 
large-scale emission.
Our \GP{} models shown here underpredict the \icecube{} results by a factor of $\sim$4, implying that hadronic sources contribute $\sim$75\% of the Galactic plane emission.

\subsection{Comparison to Gamma-Ray Observatories}

The Galactic plane large-scale \graya{} emission has been detected in the VHE energy regime by observatories such as \lhaaso{} \citep{2023PhRvL.131o1001C}.
As the $pp$ collisions create both \grays{} and neutrinos, the two emissions are connected.
From \citet{2017PTEP.2017lA105A}, the \graya{} flux can be converted to a per-flavour neutrino flux via the relation:
\begin{eqnarray}
    \frac{1}{3} \sum E_{\nu}^{2} \frac{\mathrm{d}N_{\nu}}{\mathrm{d}E_{\nu}\mathrm{d}t}(E_{\nu}) \approx \frac{1}{2} E_{\gamma}^{2} \frac{\mathrm{d}N_{\gamma}}{\mathrm{d}E_{\nu}\mathrm{d}t}(E_{\gamma}) , \label{eq:pi-to-nu conversion}
\end{eqnarray}
where the sum is over all neutrino flavours, $\nu$ and $\gamma$ subscripts denotes neutrinos or \grays{}, $E$ is the energy of the particle, $N$ is the number of particles, and $t$ is time.
The neutrino energy is given by $2E_{\nu}=E_{\gamma}$, i.e.\@ two times more energy is deposited into the \grays{} than the neutrinos.

For energies above $E_{\gamma} \approx 40$\,TeV ($E_{\nu} \approx 20$\,TeV) the pair absorption of the \grays{} on the background photon fields ($\gamma \gamma \rightarrow e^{-}e^{+}$) will reduce the \graya{} flux \citep{2006ApJ...640L.155M,2018PhRvD..98d1302P}.
Pair absorption will therefore lead to an underestimation of the expected neutrino flux from the \lhaaso{} observations.
Additionally, there will be contamination from inverse Compton (IC) emission from secondary leptons created in hadronic processes.
However, this effect is on the order of 1\% of the total emission for a purely hadronic source, and is lower for sources with a primary leptonic component.
As the IC emission from secondary leptons is negligible it is not accounted for.

Applying \autoref{eq:pi-to-nu conversion} to the \lhaaso{} \graya{} observations provides an upper limit on the expected neutrino emission.
This limit assumes no unresolved sources are present in the \lhaaso{} results and a purely hadronic origin of the diffuse emissions.
We then construct a lower limit by considering unresolved source estimates from the literature \citep[e.g.][]{2024ChPhC..48k5105C,2024NatAs...8..628Y,2025ApJ...980...17H,2025ApJ...981...93M}.
Unresolved sources are estimated to account for up to $\sim$75\% of the \lhaaso{} emission observed in \citet{2024ChPhC..48k5105C} for energies below the pair-absorption regime.
Within the pair-absorption regime the $\gamma$-to-$\nu$ conversion given in \autoref{eq:pi-to-nu conversion} is invalid.

\begin{figure}
    \centering
    \includegraphics[width=\linewidth]{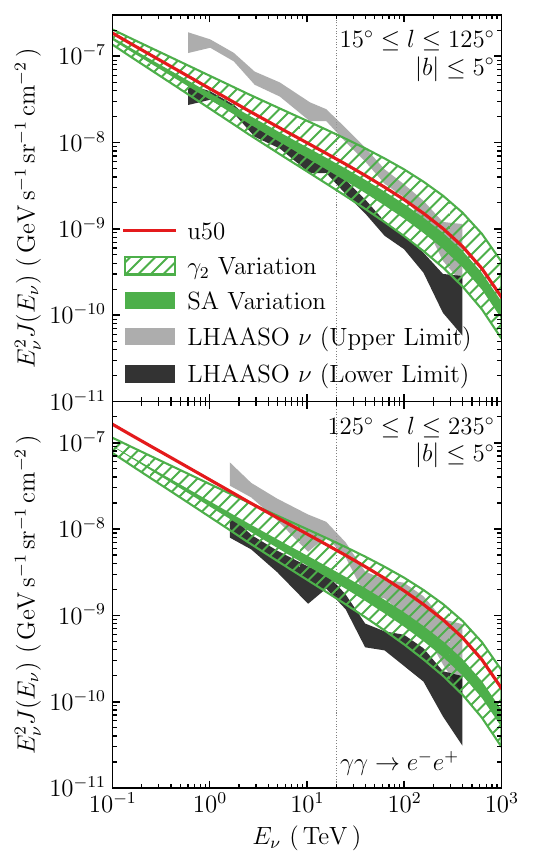}
    \caption{The \GP{} predicted diffuse per-flavour neutrino flux towards to inner (top) and outer (bottom) \lhaaso{} regions across the three source distributions (SA variation; green band) and over variations in the injection spectral index ($\gamma_{2}$ variation, green hatched band). The \lhaaso{} diffuse \graya{} results are shown after converting to expected neutrino fluxes assuming that the emission is 100\% hadronic (upper limit, grey) and 25\% hadronic (lower limit, black), with the statistical and systematic uncertainties added in quadrature. The vertical dotted line at $E_{\nu}=20$\,TeV (equivalent to $E_{\gamma}=40$\,TeV) denotes the region where pair absorption effects reduce the accuracy of the \lhaaso{} $\gamma$-to-$\nu$ conversion.}
    \label{fig:LHAASO neutrino flux}
\end{figure}

\begin{figure}
    \centering
    \includegraphics[width=\linewidth]{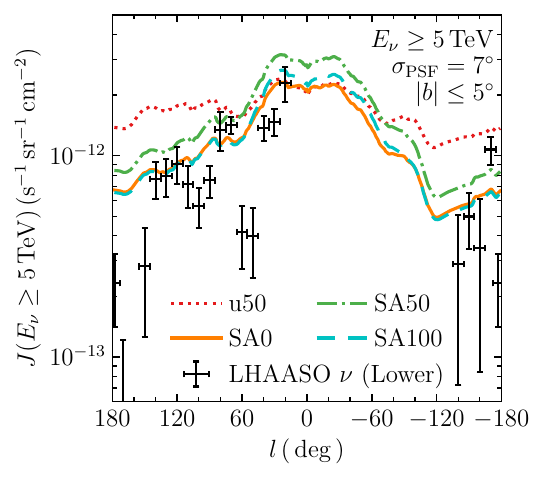}
    \caption{Longitudinal profile taken for $|b| \leq 5^{\circ}$ of the \GP{} predicted per-flavour neutrino flux at $E_{\nu}=100$\,TeV across u50 (dotted red), SA0 (solid orange), SA50 (dash-dotted green), and SA100 (dashed cyan) source distributions. The analysis was performed on the skymap with a $\sigma = 7^{\circ}$ Gaussian smear from \autoref{fig:neutrino skymaps}. Also shown is a lower-bound estimate of the \lhaaso{} data points (black) from \citep{2024ChPhC..48k5105C} after converting to a neutrino flux (\autoref{eq:pi-to-nu conversion}) and subtracting an estimated source component of 75\%.}
    \label{fig:neutrino longitudinal profile}
\end{figure}

The neutrino estimates from the \lhaaso{} \graya{} observations are compared to our \GP{} predictions in Figures \ref{fig:LHAASO neutrino flux} and \ref{fig:neutrino longitudinal profile}.
\autoref{fig:LHAASO neutrino flux} shows the upper and lower neutrino flux limits calculated from the \lhaaso{} large-scale \graya{} observations \citep{2023PhRvL.131o1001C} for both their `inner' and `outer' analysis regions.
Additionally, \autoref{fig:neutrino longitudinal profile} shows the lower neutrino flux limits calculated from the \lhaaso{} $E_{\gamma} \geq 5$\,TeV (i.e.\@ below the pair-absorption threshold) longitudinal profile from \citep{2024ChPhC..48k5105C}.
These converted \lhaaso{} results are also compared to the u50, SA0, SA50, and SA100 models after applying the \lhaaso{} source masks.
For the longitudinal profiles we also applied a $\sigma = 7^{\circ}$ Gaussian blur to our \GP{} models to matches the \icecube{} event directional uncertainty, which is also similar to the \lhaaso{} profile longitudinal bin width of 10$^{\circ}$.

From both Figures \ref{fig:LHAASO neutrino flux} and \ref{fig:neutrino longitudinal profile} we see that the diffuse \GP{} neutrino predictions are within the range of estimates computed from the \lhaaso{} \graya{} observations below the pair-absorption regime.
The spectral softening in the \lhaaso{} flux (\autoref{fig:LHAASO neutrino flux}) seen above $E_{\nu}=20$\,TeV ($E_{\gamma}=40$\,TeV) is not present in our neutrino predictions and is likely due in part to the pair absorption of the \graya{} flux above $E_{\gamma} \approx 40$\,TeV.
For the outer Galaxy we find that the \GP{} predictions lie within the upper and lower limits of the converted \lhaaso{} neutrino flux.
For the inner region, our \GP{} predictions lie on the lower limit of the converted \lhaaso{} flux, especially for the model with the softer CR injection index.

The \GP{} results from the three SA source distributions are broadly similar to one another, as expected from previous hadronic \graya{} results \citep{2023MNRAS.518.5036M}.
The spread between the SA models for $E_{\nu} \gtrsim 1$\,TeV is largely due to the minor differences in the CR spectral index between the models.
Conversely, the uniform CR density model, u50, has an increased flux of CRs in the outer Galaxy compared to the three propagation models due to the increased CR density for Galactic longitudes $|l| \geq 60$.
We find that all three SA models reproduce the general normalisation of the lower limit of the converted \lhaaso{} results for both the inner and outer regions.
The situation changes when altering the CR injection spectral index.
While the range of \GP{} results are within the uncertainties for the outer region, the softer CR index does not provide a good fit to the \lhaaso{} inner-region results.

Due to the uncertainties in the $\gamma$-to-$\nu$ flux conversion and the \lhaaso{} large-scale flux estimates we cannot confidently make conclusions on the absolute shape of the longitudinal profiles shown in \autoref{fig:neutrino longitudinal profile}.
All three SA source distributions reproduce the general shape of the \lhaaso{} observations, especially for longitudes with smaller uncertainties.
The u50 model, which is constructed similarly to the diffuse model used by the \lhaaso{} collaboration, struggles to reproduce the shape of the longitudinal profile beyond $15^{\circ} \leq |l| \leq 60^{\circ}$.
While the normalisation of the longitudinal profile depends on the spectral index of the \GP{} model, the shapes of the profiles are independent of the neutrino energy below $E_{\nu}=100$\,TeV.

\section{Discussion}

\icecube{} achieved their neutrino detection of the Galactic plane by forward-folding input models of the expected neutrino flux through the \icecube{} instrument response functions.\paperfootnote{The \icecube{} analysis uses the maximum-likelihood technique \citep{2008APh....29..299B}, with the input models being denoted $S_{i}$ in Equation~1 of the supplemental material of \citet{2023Sci...380.1338I}.}
The three input models used in the \icecube{} analysis were named $\pi^{0}$, KRA$_{\gamma}^{5}$, and KRA$_{\gamma}^{50}$.
The \icecube{} $\pi^{0}$ model is calculated from the pion-decay component of the $^{S}$S$^{Z}$4$^{R}$20$^{T}$150$^{C}$5 \GP{} model\paperfootnote{$^{S}$S$^{Z}$4$^{R}$20$^{T}$150$^{C}$5 denotes a \GP{} model that uses a source distribution based on SNRs, a 4\,kpc scale height, a radial scale distance of 20\,kpc, an atomic hydrogen spin temperature of 150\,K, and a magnitude cut applied to the dust map of $E(B-V)=5$.} from \citet{2012ApJ...750....3A} and was converted to a neutrino flux via \autoref{eq:pi-to-nu conversion}.
This 13-year-old \GP{} model was optimised for \fermi{} observations in the GeV regime, and had to be extrapolated to PeV energies \citep[see][supplemental material]{2023Sci...380.1338I}.
The KRA$_{\gamma}^{5, 50}$ models diffuse CRs using proton cut-off energies 5 and 50\,PeV, calculate the neutrino emissivity, then perform line-of-sight integrals to calculate the neutrino emission directly \citep{2015ApJ...815L..25G}.
The updates to the \GP{} framework that we use here allow the computation of the neutrino emission by performing a LoS integral over the neutrino emissivity.

While our predictions are within the observational uncertainties, our results hint towards two potentialities.
First, the fraction of unresolved sources could be a function of longitude that increases towards the GC.
Second, it is possible that the \GP{} predictions may be underestimating the hadronic emission, particularly towards the GC region.
As we have previously shown the total \GP{} \graya{} predictions provide a good description of the \lhaaso{} \graya{} observations \citep{2025ApJ...981...93M}, any potential alterations must not impact the sum of hadronic and leptonic emissions.
We find that a leptonic component is required to reproduce the \lhaaso{} results for all models, including those with harder hadronic injection spectra.
However, the large uncertainties in converting between \grays{} and neutrinos, especially in the pair-absorption regime, necessitates future neutrino observations before being utilised to constrain CR diffusion models.

\subsection{Unresolved Sources and Leptonic Components}

Due to the limited astrophysical neutrino statistics \icecube{} was unable to resolve point sources within the MW.
Hence, the \icecube{} Galactic plane neutrino flux is the sum of the diffuse emission with all that from individual sources.
The unresolved source component present in the \icecube{} results does not impact the MW detection as the diffuse emissions are spatially correlated with the VHE/UHE hadronic source emissions.
However, this source contamination implies that the number of neutrino events, which dictates the \icecube{} flux normalisation at $E_{\nu}=100$\,TeV, is overestimated for the diffuse neutrino flux.
The Galactic plane neutrino flux observation is then an upper limit on the true diffuse neutrino flux.
Estimating the fraction of emission due to unresolved sources within the MW is required to determine the accuracy of the diffuse models.

Observations from \graya{} experiments can be used to determine the unresolved source component present in the \icecube{} results.
Here we estimate the total source fraction present in the \icecube{} results we use the \lhaaso{} observations, which cover a similar energy range.
However, there are some important caveats for this comparison:
\begin{itemize}
    \item There is some fraction of leptonic emission present in \graya{} observations.
          To explain the non-detection of resolved \lhaaso{} sources in the \icecube{} results, \citet{2024JHEAp..43..140F} requires that most \lhaaso{} sources have a 20--50\% leptonic fraction, i.e.\@ there is a 50--80\% hadronic fraction.
    \item No VHE survey is complete -- all diffuse \graya{} observations have their own unresolved source components.
          Using \GP{} models, \citet{2025ApJ...980...17H} found that unresolved sources account for 5--27\% of the large-scale emission.
          However, \citet{2025JCAP...09..041V} claims that the various uncertainties in diffusion models at these energies are too large to conclude on the unresolved source fraction.
          The results from \citet{2024NatAs...8..628Y}, which are not based on CR diffusion codes, found that the unresolved emission could contribute as much as 70\%.
    \item Known \graya{} sources may not be completely masked from observations before estimating the `diffuse' emission.
          We hereafter refer to this as unmasked source emission.
          Estimates on the fraction of emission from improperly masked/unmasked sources in the \lhaaso{} results range between 0\% and 75\% of the flux depending on longitude and energy, with higher fractions found towards the GC at 10\,TeV \citep{2024ChPhC..48k5105C}.
          Our results using updated \GP{} models found that the sum of unmasked and unresolved emission accounts for 50--75\% of the \lhaaso{} large-scale emission and has a strong dependence on energy \citep{2025ApJ...981...93M}.
          We note that both estimates from \citet{2024ChPhC..48k5105C} and \citet{2025ApJ...981...93M} implicitly include a leptonic fraction.
    \item Finally, emission from the \graya{} sources resolved by \lhaaso{} will contribute to the \icecube{} observations as an unresolved neutrino component.
          The \lhaaso{} collaboration estimates the resolved source component to account for $\sim$60\% of their total large-scale emission \citep{2023PhRvL.131o1001C}.
\end{itemize}
The total source contribution to the unmasked \lhaaso{} results is then in the range of 60--90\%.
The total hadronic source fraction present within the \lhaaso{} results is then in the range of 30--72\%.
As \lhaaso{} covers the same energy range as \icecube{}, and as \icecube{} did not resolve any sources within the MW, the total hadronic fraction for \lhaaso{} should be approximately equal to the unresolved source fraction in the \icecube{} results.
In other words, we expect the unresolved source fraction in \icecube{} to be equal to 30--72\%.
Models of the diffuse neutrino emission should then under-estimate the \icecube{} Galactic plane flux by factors in the range 1.4--3.6 at $E_{\nu}=100$\,TeV.

For the three SA source distributions, our \GP{} results imply there is a source component in the \icecube{} results in the range 58--78\%, i.e.\@ we under-estimate the total Galactic plane flux by factors in the range 2.4--4.6 at 100\,TeV.
Our results lie on the lower edge of the allowed uncertainty range.
Softening the CR injection indices above 100\,GeV\,nuc$^{-1}$ underpredicts the \icecube{} neutrino flux by a factor $\sim$9 at 100\,TeV and can be excluded.
From the \icecube{} results we are unable to exclude the harder spectral model.
We also find that all of our models require some leptonic component to reproduce the \lhaaso{} observations.

The older \GP{} model utilised in the \icecube{} analysis \citep[$^{S}$S$^{Z}$4$^{R}$20$^{T}$150$^{C}$5, from][]{2012ApJ...750....3A} underpredicts the \icecube{} emission by a factor of $\sim$4.5.
Our current best-fit \GP{} models (SA0, SA50, and SA100) shown in \autoref{fig:allsky neutrino flux} underpredict the \icecube{} per-flavour neutrino emission by a factor $\sim$4.
While both the old and our updated \GP{} models are within the range of expected values, the difference to the flux reproduction is due to the improved CR flux measurements that provide a more accurate extrapolation to TeV--PeV energies compared to the older models \citep{2017ApJ...846...67P}.

Our \GP{} flux results lie on the lower limit obtained from the \icecube{} uncertainties.
While the models are within the range of factors expected from source estimates, it could hint towards a mismatch between the models and observations.
Either there is a hadronic CR source component lying on the upper edge of our expectations, there is some mismodelling, or some combination of both.
Understanding the exact dynamics requires future observations with increased neutrino statistics, as well as an analysis of the sensitivity of the \icecube{} methods to the spectral shape of the input models.

\section{Summary}

The detection of the Galactic plane in neutrinos by \icecube{} is an important step towards understanding the origin of CRs, providing a constraint on the hadronic component of the \graya{} emission.
These results have motivated the inclusion of neutrinos in the \GP{} framework.
In this work we investigated the new neutrino predictions computed from updated \GP{} models and made comparisons to the \icecube{} flux.

\icecube{} did not have enough statistics to detect the Galactic plane in a blind search, and relied on input models of the expected neutrino emission.
As no localised source component is subtracted from their results, the \icecube{} flux is an upper limit on all neutrino emission within the MW (i.e.\@ both individual sources and the diffuse emission).
The presence of this source component must be considered to prevent overfitting of diffusion simulation results to the data.
From VHE/UHE \graya{} experiments, we find that the source fraction for the \icecube{} neutrino flux is likely in the range of 30--72\%.
Diffuse neutrino emission models should then underestimate the \icecube{} total flux by factors in the range 1.4--3.6.
The updated \GP{} models investigated here are within this uncertainty range, underestimating the \icecube{} total Galactic plane neutrino flux by factors 2.4--4.6.

The significance of the \icecube{} detection of the Galactic plane depends on the input model (KRA$_{\gamma}^{50}$; 3.96$\sigma$, KRA$_{\gamma}^{5}$; 4.37$\sigma$, and \GP{}-based $\pi^{0}$; 4.71$\sigma$).
The KRA$_{\gamma}$ models have a worse spatial agreement with the observations compared to the \GP{}-based model, resulting in a lower significance for a detection of the MW.
The models presented here have an improved agreement with the \icecube{} flux normalisation and have recently been shown to agree with TeV \graya{} results from both \hess{} \citep{2023MNRAS.518.5036M} and \lhaaso{} \citep{2025ApJ...981...93M}.
Hence, we expect their use in the \icecube{} analysis pipeline will increase the neutrino detection significance for the MW.
However, strict limits on the diffuse hadronic fraction within the Galactic plane will not be possible until future observations are obtained.

For further constraints on the neutrino flux we also use the $>$10\,TeV \graya{} observations from \lhaaso{} as a proxy measurement of the neutrino emission.
We convert the \lhaaso{} large-scale \graya{} results to an upper-limit neutrino flux, and then compute the lower-limit by taking source estimates available throughout the literature.
Five of our six \GP{} models fit within this range.
Hence, we find that the current \GP{} models reproduce both the observed Galactic plane neutrino flux from \icecube{} and the estimates calculated from \lhaaso{}, within the experimental uncertainties of the observatories.
We also find that the large-scale emission results from \lhaaso{} requires some leptonic component in all cases.
We do not require any changes to our fundamental assumptions on particle injection, transport, cross sections, the ISM distributions, or the inclusion of dark matter, to reproduce the current observations of the Galactic plane \graya{} or neutrino emission.

\begin{acknowledgments}
\GP{} development is partially funded via NASA grants 80NSSC22K0477, 80NSSC22K0718, and 80NSSC23K0169.
CR data was compiled via the Cosmic-Ray Data Base (CRDB) \citep{2014A&A...569A..32M,2020Univ....6..102M,2023EPJC...83..971M}.
We thank S.~Sclafani for guidance on interpreting the \icecube{} results and K.~Fang for useful discussions.
\end{acknowledgments}

\begin{contribution}



PDM was responsible for the analysis and writing of the manuscript.
TAP, IVM obtained the funding and edited the manuscript.
GJ, GPR edited the manuscript.
  
\end{contribution}

%

\software{astropy \citep{2013A&A...558A..33A, 2018AJ....156..123A},
          \GP{} \citep{1998ApJ...493..694M,1998ApJ...509..212S,2022ApJS..262...30P},
          HEALPix \citep{2005ApJ...622..759G},
          matplotlib \citep{Hunter:2007},
          numpy \citep{harris2020array}, and
          scipy \citep{2020SciPy-NMeth}.}



\section{Appendix} \label{sect:appendix}

\begin{figure}
    \centering
    \includegraphics[width=\linewidth]{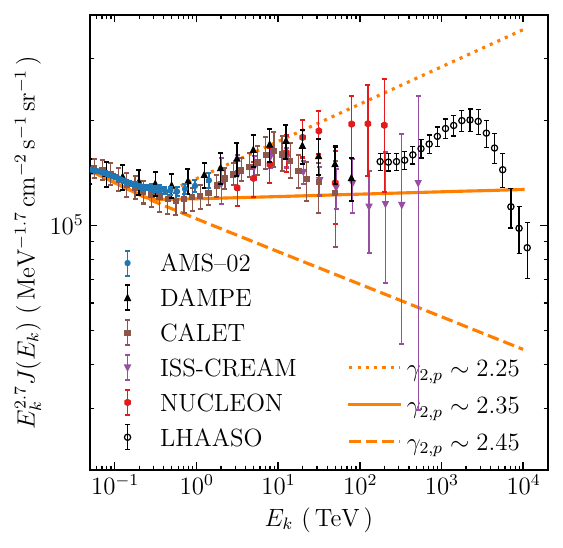}
    \caption{Proton kinetic energy spectrum above 50\,GeV at the Solar location. The \GP{} propagated spectrum is shown for three values of the highest-energy injection index. Observational datapoints are from AMS--02 \citep[blue circles;][]{2021PhR...894....1A}, DAMPE \citep[black triangles;][]{2019SciA....5.3793A}, CALET \citep[brown squares;][]{2022PhRvL.129j1102A}, ISS--CREAM \citep[purple triangles;][]{2022ApJ...940..107C}, NUCLEON \citep[red hexagons;][]{2019AdSpR..64.2546G}, and LHAASO \citep[black circles;][]{2025arXiv250514447T}.}
    \label{fig:proton flux}
\end{figure}

\begin{figure}
    \centering
    \includegraphics[width=\linewidth]{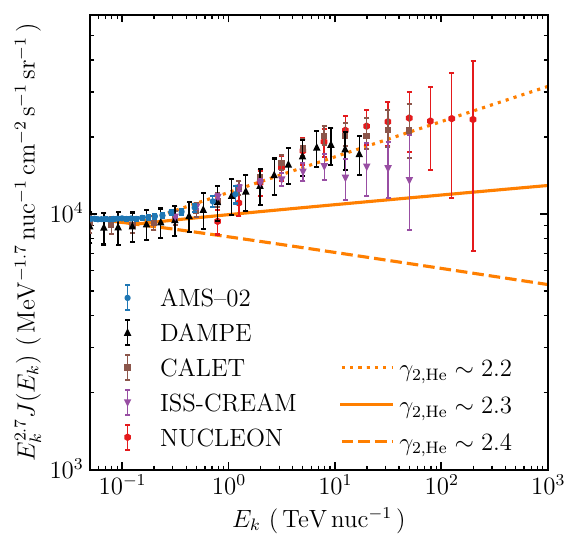}
    \caption{Helium kinetic energy spectrum above 50\,GeV at the Solar location. The \GP{} propagated spectrum is shown for three values of the highest-energy injection index. Observational datapoints are from AMS--02 \citep[blue circles;][]{2021PhR...894....1A}, DAMPE \citep[black triangles;][]{2021PhRvL.126t1102A}, CALET \citep[brown squares;][]{2023PhRvL.130q1002A}, CREAM--I+III \citep[purple triangles;][]{2017ApJ...839....5Y}, and NUCLEON \citep[red hexagons;][]{2019AdSpR..64.2546G}.}
    \label{fig:helium flux}
\end{figure}

We fit our CR injection spectra such that the propagated flux agrees with local observations at 100\,GeV, and extrapolate to higher energies \citep[see][]{2023MNRAS.518.5036M}.
Given the uncertainties in the $>$10\,TeV regime, we test a range of spectral indices for the injected CRs for rigidities $R>266$\,GV.
The spectral indices are denoted $\gamma_{2,p}$ and $\gamma_{2,\mathrm{He}}$ for protons and helium, respectively.
We show the \GP{} propagated CR proton and helium spectra for the SA50 source distribution at the Solar location along with local data in Figures \ref{fig:proton flux} and \ref{fig:helium flux}.

\bibliography{ref.bib}{}
\bibliographystyle{aasjournal}



\end{document}